\documentclass[aps,pra,reprint,amsmath,twocolumn,showpacs,superscriptaddress]{revtex4-1}  
\usepackage{graphicx}  
\usepackage{dcolumn}   
\usepackage{amssymb}   

\usepackage{color}
\usepackage[usenames,dvipsnames,svgnames]{xcolor}
\usepackage[utf8]{inputenc}
\usepackage{tabularx}
\usepackage{ragged2e}
\allowdisplaybreaks
\usepackage{amsmath}
\usepackage{bm} 
\usepackage{mathtools}
\usepackage{afterpage}

\DeclarePairedDelimiter\abs{\lvert}{\rvert}
\newcommand{\difdisp}[2]{\frac{\mathrm{d}#1}{\mathrm{d}#2}} 
\newcommand{\pdifdisp}[2]{\frac{\partial #1}{\partial #2}} 
 
\newcommand{\imi}[0]{\mathrm{i}} 
\newcommand{\difd}[0]{\mathrm{d}} %

\newcommand{\floor}[1]{\lfloor #1 \rfloor} 

 \DeclareFontFamily{U}{wncy}{}
    \DeclareFontShape{U}{wncy}{m}{n}{<->wncyr10}{}
    \DeclareSymbolFont{mcy}{U}{wncy}{m}{n}
    \DeclareMathSymbol{\Sh}{\mathord}{mcy}{"58} 

\begin{document}



\title{Self-synchronization Phenomena in the Lugiato-Lefever Equation}

\author{Hossein Taheri}
\email{h.taheri@gatech.edu}
\affiliation{School of Electrical and Computer Engineering, Georgia Institute of Technology, Atlanta GA, USA}
\author{Pascal Del'Haye}
\affiliation{National Physical Laboratory, Teddington, UK}
\author{Ali A. Eftekhar}
\affiliation{School of Electrical and Computer Engineering, Georgia Institute of Technology, Atlanta GA, USA}
\author{Kurt Wiesenfeld}
\email{kurt.wiesenfeld@physics.gatech.edu}
\affiliation{Center for Nonlinear Science, School of Physics, Georgia Institute of Technology, Atlanta GA, USA}
\author{Ali Adibi}
\email{ali.adibi@ece.gatech.edu}
\affiliation{School of Electrical and Computer Engineering, Georgia Institute of Technology, Atlanta GA, USA}


\begin{abstract}
The damped driven nonlinear Schr\"{o}dinger equation (NLSE) has been used to understand a range of physical phenomena in diverse systems. Studying this equation in the context of optical hyper-parametric oscillators in anomalous-dispersion dissipative cavities, where NLSE is usually referred to as the Lugiato-Lefever equation (LLE), we are led to a new, reduced nonlinear oscillator model which uncovers the essence of the spontaneous creation of sharply peaked pulses in optical resonators. We identify attracting solutions for this model which correspond to stable cavity solitons and Turing patterns, and study their degree of stability. The reduced model embodies the fundamental connection between mode synchronization and spatiotemporal pattern formation, and represents a novel class of self-synchronization processes in which coupling between nonlinear oscillators is governed by energy and momentum conservation.


\end{abstract}

\maketitle

\section{Introduction}\label{sec:introduction}
Self-organization is an intriguing aspect of many nonlinear systems far from equilibrium, which leads to the emergence of coherent spatiotemporal structures \cite{haken2004synergetics}. Many such nonlinear systems have been modeled by the externally driven damped nonlinear Schr\"{o}dinger equation (NLSE). Examples include such diverse systems as Josephson junctions, charge-density-waves, quantum Hall ferromagnets and ferromagnets in microwave fields, RF-driven plasmas, shear flows in liquid crystals, and atmospheric and ocean waves \cite{[{See~}][{ and references [1-18] therein.}]barashenkov2011travelling}. The NLSE admits spatiotemporal sharply peaked solutions (e.g., dissipative solitons), but a central mystery remains: while it is understood that such solutions occur because of \emph{phase locking}, no formal model is currently available to explain the underlying self-synchronization mechanism. In this paper, we introduce a reduced phase model which captures the fundamental connection between mode synchronization and pulse formation. While the results presented here are generic from the mathematical perspective, considering them within a specific physical system allows a more lucid presentation and interpretation of the results.  Consequently, we consider the damped driven NLSE in the context of frequency combs based on dissipative optical cavities \cite{akhmediev2005dissipative, lugiato1987spatial, kippenberg2011microresonatorbased}.\newline
\indent A high-Q (quality-factor) optical resonator made of Kerr-nonlinear material and pumped by a continuous wave (CW) laser forms a hyper-parametric oscillator based on nonlinear four-wave mixing (FWM) \cite{del2007monolithic, savchenkov2008tunable}, and can generate an optical frequency comb: an array of frequencies spaced by (an integer multiple of) the resonator free spectral range (FSR). The generation of a frequency comb with equidistant teeth is, however, not enough; temporal pulse generation requires also the mutual phase locking (synchronized oscillation) of the frequency comb teeth. Unlike pulsed lasers, pulsation in dissipative optical resonators requires neither active nor passive mode locking elements (e.g. modulators or saturable absorbers) \cite{haus2000mode, kippenberg2011microresonatorbased}. Rather, pulsed states arise naturally from a simple damped driven NLSE, in this context commonly called the Lugiato-Lefever equation (LLE) \cite{lugiato1987spatial, haelterman1992dissipative, matsko2011mode, barashenkov2011travelling, coen2013modeling, chembo2013spatiotemporal, herr2014temporal}. Two categories of stable pulsed solutions have been identified for the LLE: stable modulation instability (also called hyper-parametric oscillations or Turing rolls) and stable cavity solitons \cite{barashenkov1991stability, matsko2011mode, matsko2012hard, coen2013modeling, chembo2013spatiotemporal, chembo2014stability}. Owing to their stability, these phase-locked combs have been used to demonstrate chip-scale low-phase-noise radio frequency sources \cite{liang2015high} and high-speed coherent communication \cite{pfeifle2015optimally, pfeifle2014coherent}.\newline
\indent Phase locking in optical microresonators has been studied in terms of the cascaded emergence of phase-locked triplets \cite{coillet2014robustness} and injection locking of overlapping comb bunches \cite{del2014self}. Additionally, few-mode models have explained the phase offset between the pumped mode and the rest of the comb teeth \cite{loh2014phase,taheri2015anatomy}, and have shed light on the temporal evolution of comb harmonic phases \cite{taheri2017anatomy}. More recently, Wen \emph{et al.}\ \cite{wen2014self} have emphasized the link between oscillator synchronization---most famously described by the Kuramoto model---and the onset of pulsing behavior. However, while stable ultrashort pulses have been demonstrated in a variety of platforms \cite{savchenkov2008tunable, herr2014temporal, saha2013modelocking, vahala2015soliton}, their underlying phase locking mechanism is still unknown. The reduced model introduced in this paper reveals the underlying nonlinear interactions responsible for the spontaneous creation of pulses in optical resonators with anomalous dispersion. The modal interactions in the LLE are the result of the cubic (Kerr) nonlinearity and their specific form reflects conservation of energy and momentum. Consequently, the phase couplings in our model are \emph{ternary} (i.e., they involve three-variable combinations) rather than binary, as in typical phase models \cite{pikovsky2003synchronization}. Our model admits attracting solutions which correspond to stable cavity solitons and Turing rolls. We show that the phase stability of steady-state LLE solutions in the strong pumping regime can be studied easily using this model. Moreover, our model sheds light on the role of MI and chaos in the generation and stability of Turing rolls and solitons.\newline

\section{Reduction of the Lugiato-Lefever Equation}\label{sec:reduction}
\noindent Creation of sharply peaked solutions in dissipative optical cavities relies on the establishment of a fine balance between nonlinearity, dispersion (or, in the case of spatial cavities, diffraction), parametric gain, and cavity loss \cite{grelu2012dissipative}. The dynamics of this complex interaction is described by the LLE, which is a nonlinear partial differential equation with periodic boundary conditions for the intra-cavity field envelope in a slow and a fast time variable \cite{haelterman1992dissipative,coen2013modeling} or, equivalently, in time and the azimuthal angle around the whispering-gallery-mode resonator \cite{chembo2013spatiotemporal}. The cubic nonlinear term in the LLE leads to a rich interplay between the power and phase dynamics of the comb teeth. To uncover the self-synchronization mechanism leading to phase locking in this equation, as will become clear in this section, we make experimentally-motivated assumptions on the power spectrum. This simplification allows us to separate the evolution of the power spectrum from that of the phase and arrive at a reduced model (a phase model) which embodies the fundamental phase locking mechanism enabled by the nonlinearity in the LLE.
%
\begin{figure*}[tbp]
\centering
\includegraphics[width=0.85\textwidth]{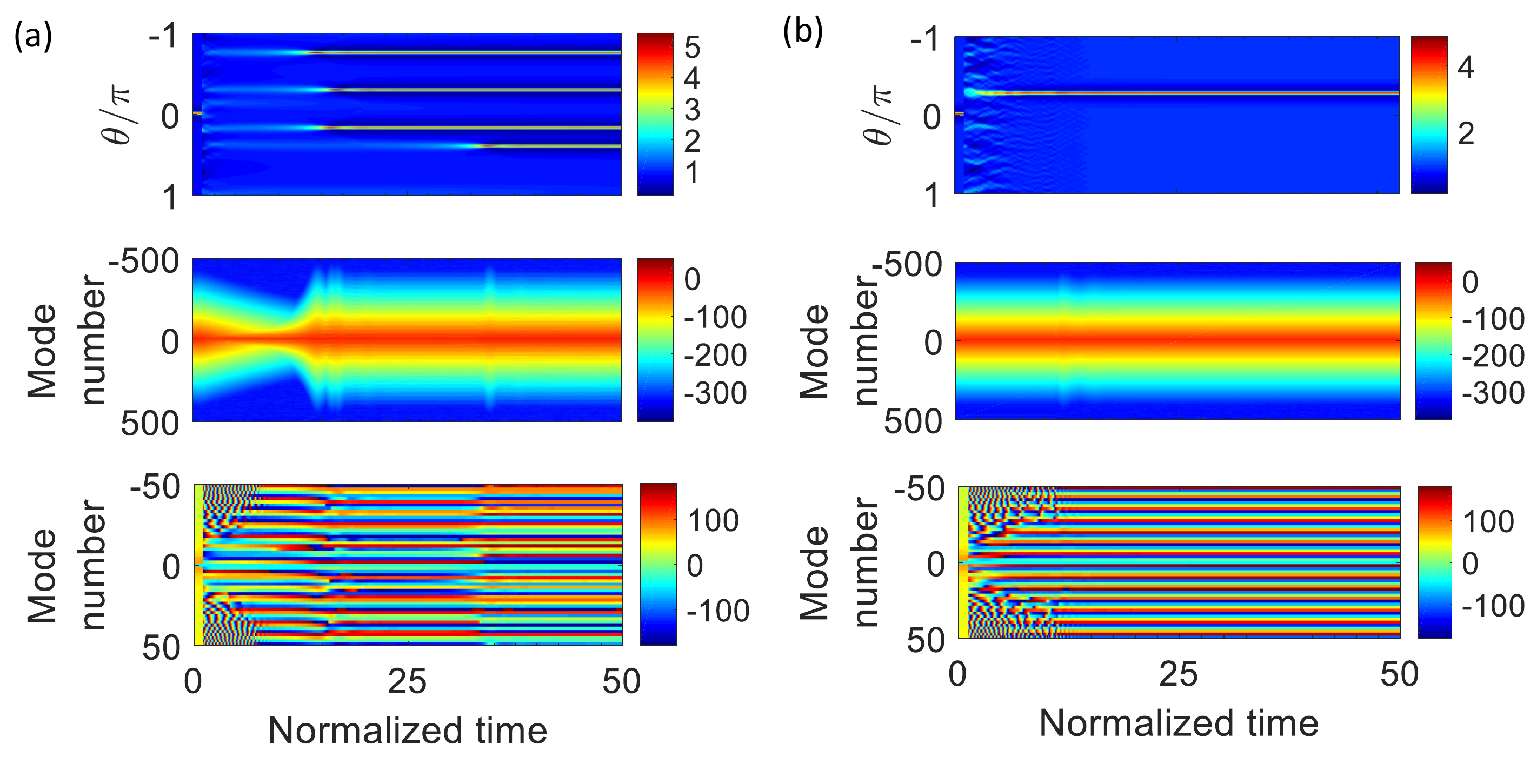}
\caption{\label{fig:prelim:synchtest}Phase locking after randomizing the phase profile of a single-soliton LLE solution. A dissipative soliton is propagated in time under the LLE. The soliton phase profile is suddenly randomized at $\tau = 1$. (a) Integration of the LLE is continued normally after phase randomization. As a result of the interplay between power and phase evolution, the single-soliton state is lost but multiple solitons may appear, pointing at the existence of a phase locking mechanism. (b) To suppress the influence of comb power evolution on the phase recovery, the power spectrum of a single soliton is enforced in every integration step after phase randomization, and the single soliton peak is seen to be recovered. The upper panels illustrate the temporal evolution of the intra-cavity waveform by color density, while the middle and lower panels shows the frequency comb power and phase profiles, respectively.}
\end{figure*}
%

In normalized form, the LLE reads
\begin{equation}\label{eq:LLE}
\pdifdisp{\psi}{\tau}=-(1+\imi\alpha)\psi-\imi \frac{d_2}{2} \frac{\partial^2\psi}{\partial\theta^2}+\imi|\psi|^2\psi+F,
\end{equation}
where $\psi(\theta, \tau)$ and $F(\theta, \tau)$ are the field envelope and pump amplitude respectively, both normalized to the sideband generation threshold, $\alpha$ and $d_2$ are the pump-resonance detuning and second-order dispersion coefficient, each normalized to the half-linewidth of the pumped resonance, and $\tau$ is the time normalized to half of the cavity photon lifetime \cite{chembo2013spatiotemporal} ($d_2 < 0$ for anomalous dispersion). As noted earlier, the LLE has stable dissipative soliton and Turing roll solutions. If a phase locking mechanism exists in the LLE, when one of its phase-locked steady-state solutions, e.g.~a single soliton, is used as initial condition for propagation with time and its phase spectrum is randomized, we would expect the phase locking mechanism to recover the soliton phase after some time; see Fig.~\ref{fig:prelim:synchtest}. Because of the interplay of the power and phase dynamics, more than one local peak may appear after randomizing the phase profile as seen in Fig.~\ref{fig:prelim:synchtest}(a). To appreciate the influence of separating power and phase dynamics, it is possible to enforce the power spectrum of a single-soliton solution in every step of integration of the LLE when propagating the solution in time. Then, the system converges to the simpler phase-locked state of a single soliton, as can be seen in Fig.~\ref{fig:prelim:synchtest}(b). The difference between the smooth (before randomizing the phases) and striped (after pulse recovery) phase profile (lower panel) in Fig.~\ref{fig:prelim:synchtest}(b) stems from the linear added phase due to the shift of the recovered soliton peak and wrapping of the phase between $-\pi$ and $\pi$. The slope of the linear phase profile, as we will show, depends on the initial random phase profile when the locking process kicks in and its arbitrary character is a result of the rotational symmetry of the resonator; see the discussion about the zero eigenvalue in Section \ref{sec:stability}. The phase profile (lower panel) of the recovered multi-soliton state of Fig.~\ref{fig:prelim:synchtest}(a) is constant with time after the fourth peak appears but, in contrast to the single-soliton phase profile of the lower panel in Fig.~\ref{fig:prelim:synchtest}(b), does not have a regular pattern repeating with the mode number. It is worth noting that we have used a very extreme phase randomization in Fig.~\ref{fig:prelim:synchtest}, i.e., random phases chosen from a uniform distribution over $(-\pi, \pi ]$. If, instead, a normal distribution with standard deviation equal to a fraction of the period (e.g., $\pi/4$) is used, a single soliton, rather than multiple solitons, is more likely to be recovered even without enforcing the single soliton power spectrum.
%
\begin{figure}[tbp]
\centering
\includegraphics[width=0.45\textwidth]{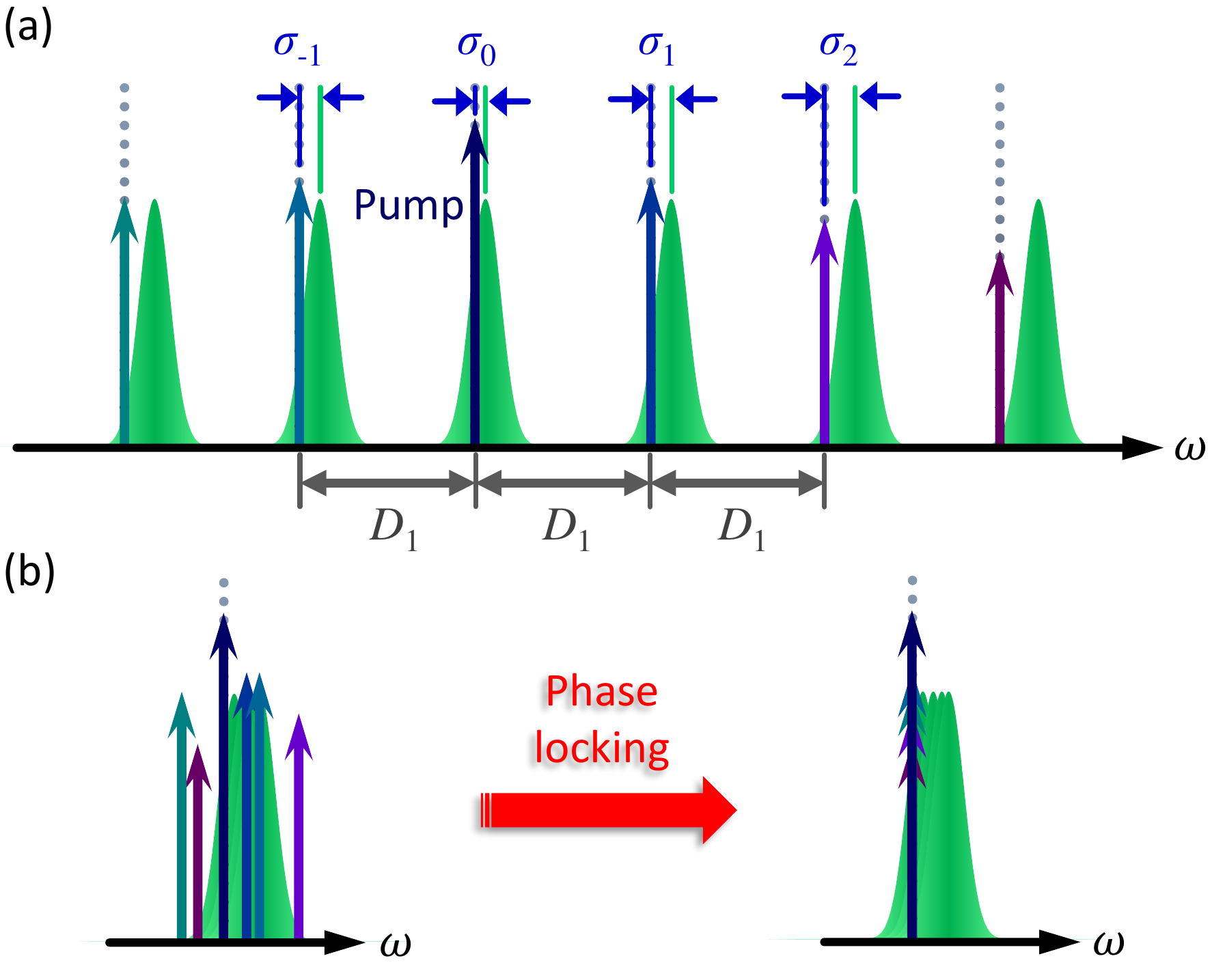}
\caption{\label{fig:prelim:lockconcept}Frequency-domain interpretation of the LLE. The LLE, Eq.~(\ref{eq:LLE}), defines a grid (dotted lines) with spacing equal to the resonator FSR at the pump frequency, $D_1$, and is written in a rotating reference frame, \cite{chembo2013spatiotemporal}. In the frequency domain, moving to the rotating frame translates into removing the spacing between the grid sites by folding the schematic of panel (a) such that all the dotted lines coincide. Because of modal dispersion, the modal resonances (green) will not all fall at the same position. Before phase locking, the different comb teeth may be at any spectral position around their corresponding resonance. Phase locking is established when all of the comb teeth align with the pump and, additionally, oscillate in synchrony. $\sigma_\eta \propto \alpha_\eta$ is the dimensional detuning of comb tooth $\eta$ from its nearest resonance.}
\end{figure}
%

Figure~\ref{fig:prelim:synchtest} suggests that a phase-locking mechanism does indeed underlie pulse formation in the LLE. To understand this mechanism, we consider comb generation in the frequency domain. The discrete-time Fourier transform of Eq.~(\ref{eq:LLE}) (with the azimuthal angle $\theta$ and comb mode number $\eta$ as conjugate variables \cite{[{For the Fourier pairs we have used the following equations and sign convention:~$$\tilde{a}_\eta(\tau)=\frac{1}{2\pi}\int_{-\pi}^{\pi}\difd\theta \, \psi(\theta, \tau) \, \exp(-\imi\eta\theta),$$ and $$\psi(\theta, \tau)=\sum_{\eta=-N}^N \tilde{a}_\eta(\tau) \,  \exp(+\imi\eta\theta),$$ where, $N$ is an integer. In the formal definition of the discrete-time Fourier transform, $N$ is replaced with $\infty$. For functions of interest to this work, the comb span $N$, although possibly large (e.g., a few thousands), is finite. See Sec. 2.7 of~}] oppenheim1989dsp}), yields an equivalent set of coupled nonlinear ordinary differential equations (ODEs) \cite{chembo2010modal},
\begin{equation}\label{eq:CNODE}
\difdisp{\tilde{a}_\eta}{\tau}=-(1+\imi\alpha_\eta)\tilde{a}_\eta+\imi\sum_{l,\, m,\, n}\tilde{a}_l \tilde{a}_m^*\tilde{a}_n \, \delta_{\eta_{lmn}\eta}+\tilde{F}_\eta,
\end{equation}
for the temporal evolution of the complex comb teeth amplitudes $\tilde{a}_\eta$ (with magnitude $|\tilde{a}_\eta|=a_\eta$ and phase $\angle{\tilde{a}_\eta}=\phi_\eta$) which make up the spatiotemporal field envelope through $\psi(\theta, \tau)=\sum\nolimits_{\eta=-N}^N \tilde{a}_\eta \exp(\imi\eta\theta)$. In this picture, each comb mode is a nonlinear oscillator and one of the coupled ODEs follows the temporal evolution of its complex amplitude. In Eq.~(\ref{eq:CNODE}), $\alpha_\eta = \alpha-d_2 \eta^2 / 2$ is the detuning of comb tooth $\eta$ from its neighboring resonance, $\delta_{pq}$ (for integers $p$ and $q$) is the Kronecker delta, $\eta_{lmn}=l-m+n$, and $l$, $m$, and $n$ are integers; modes are numbered relative to the pumped mode for which $\eta = 0$. We consider CW pumping for which $\tilde{F}_\eta=\delta_{0\eta}F_{\mathrm{P}}\exp(\imi\phi_{\mathrm{P}})$, $F_\mathrm{P}$ being proportional to the pump magnitude and $\phi_{\mathrm{P}}$ representing its phase. \newline
\indent The LLE defines a grid in the frequency domain (dotted lines in Fig.~\ref{fig:prelim:lockconcept}) where the spacing between the grid sites is equal to the resonator FSR ($D_1$, see Appendix~\ref{app:derivation} and \cite{chembo2013spatiotemporal}) at the pumped mode. The standard LLE \cite{chembo2013spatiotemporal} is written in a rotating reference frame such that in its derivation a term $D_1 \partial \psi / \partial \theta$ is removed from the equation to yield Eq.~(\ref{eq:LLE}). In the frequency domain, this change translates into removing a term $\imi \eta D_1$ from each of the coupled ODEs with $\eta \ne 0$, which, in turn, amounts to removing the spacing between the grid sites by folding Fig.~\ref{fig:prelim:lockconcept}(a) such that all the dotted lines coincide. Because of the resonator modal dispersion, the modal resonances (green) will not all fall at the same position. Before phase locking, the different comb harmonics (teeth) may be at any spectral position around their corresponding resonance. Phase locking is established when all of the comb lines align with the pump and, additionally, oscillate synchronously. The phase profile $\phi_\eta$ of the comb teeth complex amplitudes $\tilde{a}_\eta$ in Eq.~(\ref{eq:CNODE}) captures both the alignment and the synchronized oscillation of the comb teeth.\newline
\indent Experimentally, Turing rolls arise from the intra-cavity equilibrium field through modulation instability of vacuum fluctuations and correspond, in the frequency domain, to combs that usually have multiple-FSR spacing between their adjacent teeth. Solitons, on the other hand, are coherent combs with single-FSR spacing. Experimental and theoretical studies have suggested that solitons are not accessible from the CW intra-cavity field without seeding \cite{taheri2015soliton, jang2015writing}, changing the pump frequency or power \cite{matsko2012hard, lamont2013route, herr2014temporal,jaramillo2015deterministic}, or a suitable input pulse \cite{leo2010temporal}. In the model introduced here, we treat solitons and rolls in a unified manner. For solitons, $\mathord{\eta\in\{0, \pm 1, \dots, \pm N\}}$ while for rolls $\mathord{\eta\in\{0, \pm\mu, \pm2\mu, \dots, \pm N\mu\}}$, where $N$ is a positive integer and the integer $\mu \geq 1$ is the mode number at which MI gain peaks (the first pump sidebands are generated) \cite{chembo2010spectrum, chembo2014stability, taheri2017anatomy}. \newline
\begin{figure*}[tbp]
\centering
\includegraphics[width=0.7\textwidth]{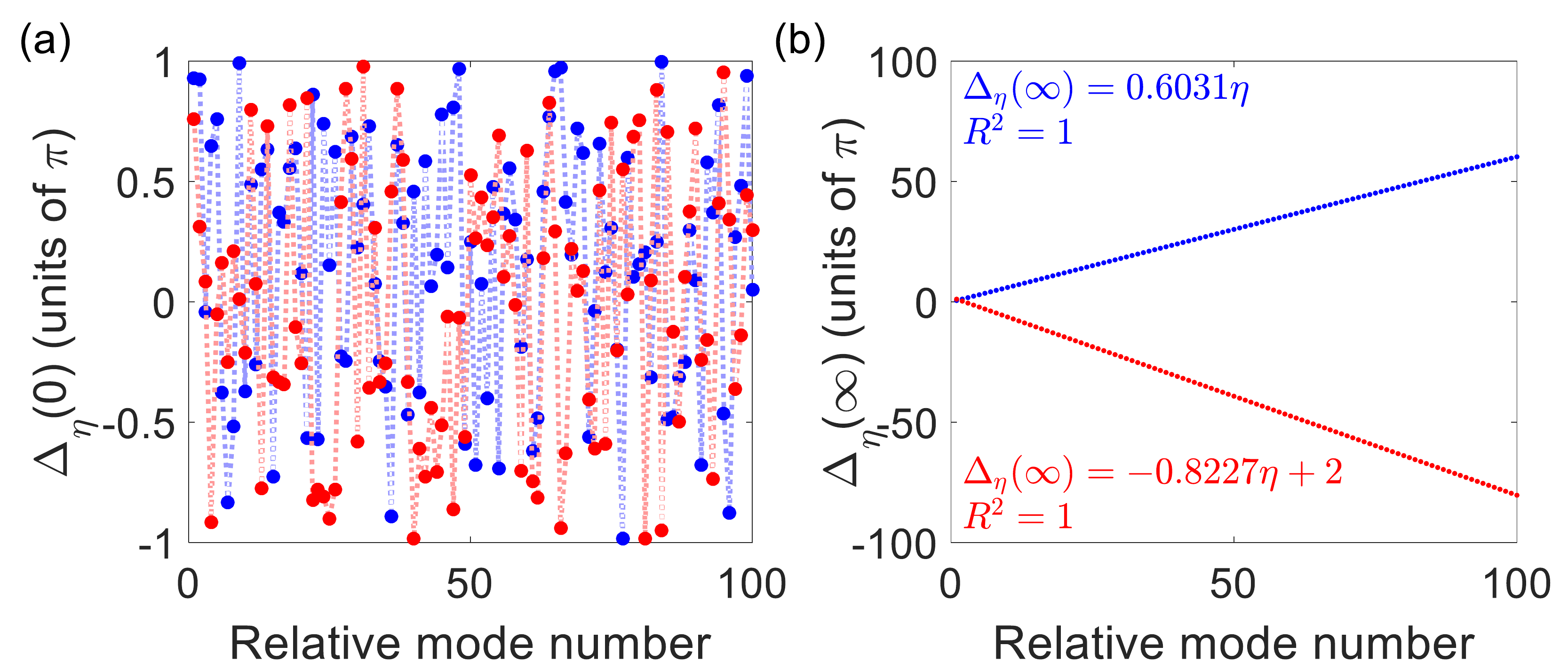}
\caption{\label{fig:prelim:numsol}(Color online) Numerical solutions of Eq.~(\ref{eq:pl}). (a) Two examples of the phase differences (PDs) at the onset of integration (initial conditions). These initial PDs have uniform probability density over $(-\pi, \pi]$. (b) Steady-state PDs for the initial conditions shown in (a). It is seen that the final PDs lie on straight lines irrespective of the initial conditions, but different initial values lead to different slopes for these lines. The upper blue (dark gray) line corresponds to the blue (dark gray) initial values, while the lower red (light gray) line corresponds to the red (light gray) initial values in (a). We verify the linearity of the final PDs through fitting a straight line to them (equations on the plot) and calculating the coefficient of determination (R-squared). $R^2=1$ shows that the PDs do indeed lie on straight lines.}
\end{figure*}
\indent Experiments and numerical simulations suggest that for stable solutions, the power of the pumped mode is much larger than the other modes (the strong pumping regime) and that in the absence of third- and higher-order dispersion \cite{taheri2017hod}, the power spectrum of these solutions is symmetric with respect to the pumped mode \cite{saha2013modelocking, herr2014temporal, chembo2014stability} (see, e.g., the inset curves $a_\eta^2$ vs.\ mode number in Fig.~\ref{fig:prelim:phasealigned}). Therefore, we exploit the symmetry of the power spectrum, adopt a perturbative approach (with $a_\eta$ for $\eta\ne0$ as the small parameters), and retain terms with at least one contribution from the pumped mode $a_0$ in the triple summations in Eq.~(\ref{eq:CNODE}). Equations of motion for the magnitudes $a_\eta$ and phases $\phi_\eta$ can readily be found by using $\tilde{a}_\eta=a_\eta \exp(\imi\phi_\eta)$ in the resulting truncated equations, dividing by $\exp(\imi\phi_\eta)$, and separating the real and imaginary parts (see Appendix~\ref{app:derivation} for details). Our approach follows that of Ref.~\cite{wen2014self}, with the generalization that here the comb teeth magnitudes are not required to be equal. \newline
\indent The magnitude and phase equations for the pumped mode include no linear contributions from $a_{\eta\ne0}$ and read
\begin{subequations}
\begin{align}
\difdisp{}{\tau} \ln(a_0)=\frac{F_{\mathrm{P}}}{a_0} \cos(\phi_\mathrm{P}-\phi_0)-1, \label{eq:pumpedamp} \\
\dot{\phi}_0 = \frac{F_{\mathrm{P}}}{a_0} \sin(\phi_\mathrm{P}-\phi_0)-\alpha+a_0^2. \label{eq:pumpedphase}
\end{align} 
\end{subequations}
The solutions settle on a fast time scale to the equilibrium intra-cavity field $\psi_\mathrm{e}=a_0\exp(\imi\phi_0)$ \cite{taheri2017anatomy}; subsequently, $a_0$ and $\phi_0$ can be treated as constants to first order in $a_{\eta\ne0}$.\newline
\indent Equations of motion for the centered phase averages $\zeta_\eta=\bar{\phi}_\eta-\phi_0$, where the phase average $\bar{\phi}_\eta=(\phi_\eta+\phi_{-\eta})/2$ is centered to the pumped mode phase $\phi_0$, can be found using the phase equations for $\phi_{-\eta}$, $\phi_{+\eta}$, and $\phi_0$. This equation, to lowest non-zero order in $a_{\eta\ne 0}$, takes the form
\begin{align}\label{eq:avetrunc}
\difdisp{}{\tau}\zeta_\eta = \frac{1}{2}d_2\eta^2 + a_0^2[1+\cos(2\zeta_\eta)]-\frac{F_{\mathrm{P}}}{a_0}\sin(\phi_{\mathrm{P}}-\phi_0),
\end{align}
and can be integrated directly to give
\begin{equation}\label{eq:antisym}
\tan{\zeta_\eta}=\sqrt{\abs*{\frac{C+2}{C}}}\tanh[\sqrt{|C(C+2)|}a_0^2(\tau-\tau_0)].
\end{equation}
Here $C=d_2\eta^2/2a_0^2-F_\mathrm{P}\sin(\phi_\mathrm{P}-\phi_0)/a_0^3$, and $\tau_0$ accounts for constants of integration (or initial conditions). Equation~(\ref{eq:antisym}) holds when $|2a_0^2-\alpha+d_2\eta^2/2|<a_0^2$, a condition that is automatically satisfied when MI gain exists (see Appendix~\ref{app:derivation}).  Because the hyperbolic function approaches unity as $\mathord{\tau\to\infty}$, $\bar{\phi}_\eta$ reaches the same constant irrespective of the initial conditions.  Since $\bar{\phi}_\eta$ is fixed, each pair of phases $\phi_{\pm \eta}$ must take values symmetrically located relative to the constant average. We will refer to this as phase ``\emph{anti-symmetrization}'', following the terminology of \cite{wen2014self}.  Once established, phase anti-symmetrization means each centered phase average $\zeta_\eta$ can be treated as a constant to first order in $a_{\eta\ne0}$. \newline
\indent The equations of motion for the phase differences (PDs) defined by $\Delta_\eta=(\phi_\eta-\phi_{-\eta})/2$, 
\begin{equation}\label{eq:pl}
\difdisp{\Delta_\eta}{\tau}=a_0\sum\nolimits_l K(l,\eta) \sin(\Delta_l+\Delta_{\eta-l}-\Delta_\eta),
\end{equation}
are found by combining the phase dynamics equations for each $\pm\eta$ mode pair (see Appendix \ref{app:derivation}). Here, $K(l,\eta)=a_\eta^{-1}a_l a_{\eta-l} \{2 \sin(\zeta_\eta-\zeta_{\eta-l}+\zeta_l)+\sin(\zeta_\eta-\zeta_{\eta-l}-\zeta_l)\}$ is the coupling coefficient for the pump--non-degenerate interaction of comb teeth labeled $0$, $l$, $\eta-l$, and $\eta$. Equation (\ref{eq:pl}) shows that the particular value of the pumped mode power $a_0^2$ only amounts to a re-scaling of time. This set of equations is the model which governs the long time evolution of phases in the system, and in particular provides insight as to how it displays spatiotemporal pulse formation. On the one hand, it is a {\it phase model}, and in this sense is a member of a familiar family of models, like the Adler equation \cite{adler1946locking} or the Kuramoto model \cite{strogatz2000kuramoto}, used to study spontaneous synchronization. On the other hand, Eq.(\ref{eq:pl}) is unfamiliar, involving ternary phase interactions rather than binary ones. In the remainder of this paper, we will study solutions of this reduced phase model, compare them with solutions of the LLE, and analyze their stability.\newline
\section{Reduced Equation Fixed Points}\label{sec:fixed_points}
\noindent It can readily be verified, through direct substitution, that a family of fixed point solutions of Eq.~(\ref{eq:pl}) is $\Delta_\eta=s_0\eta+k\pi$, where $s_0$ is an arbitrary constant and $k$ is an integer.  These solutions imply that the phases have aligned: the slope of the line passing through the phases of any pair of comb teeth $\eta$ and $-\eta$ will be the same and equal to $s_0$, i.e., $(\phi_\eta-\phi_{-\eta})/2\eta=s_0$.\newline
\begin{figure*}[tbp]
\centering
\includegraphics[width=0.7\textwidth]{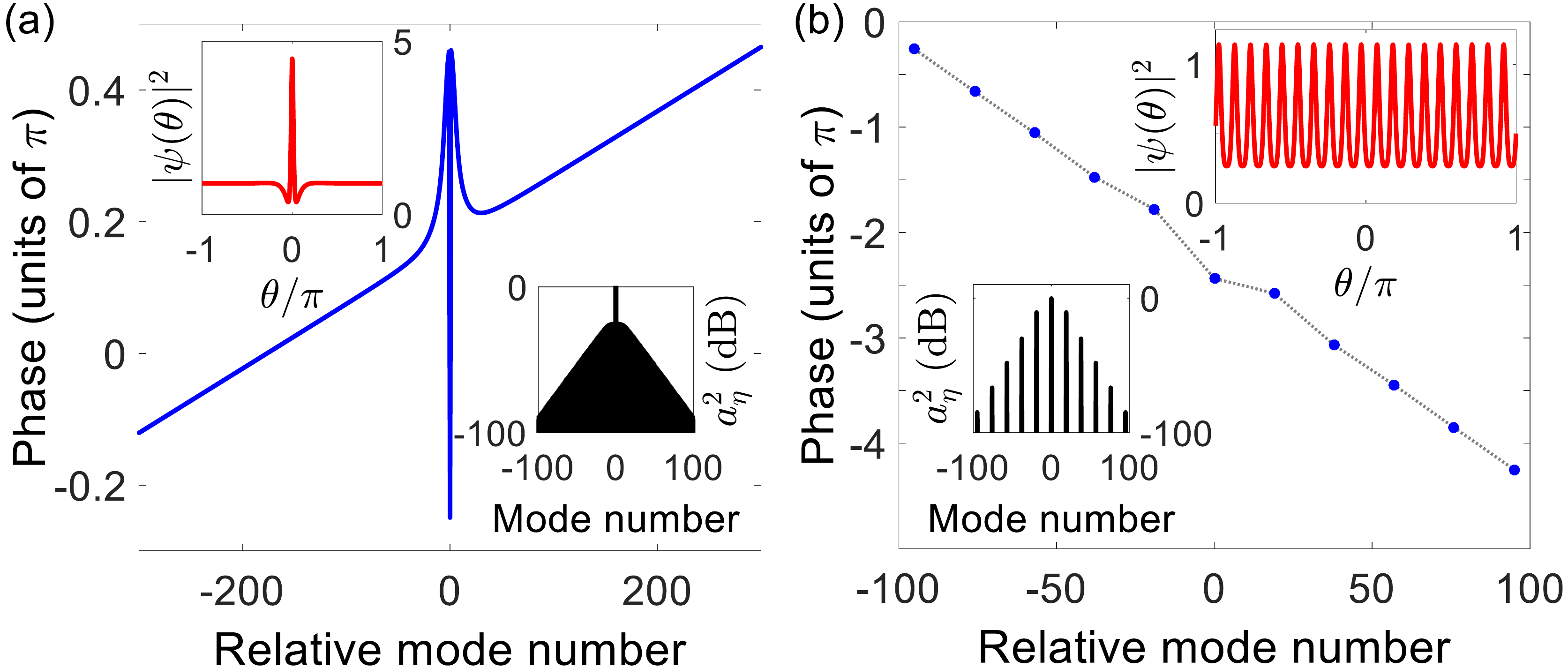}
\caption{\label{fig:prelim:phasealigned}Phase alignment in (a) solitons and (b) Turing rolls in the steady-state solutions of Eqs.~(\ref{eq:LLE}) and (\ref{eq:CNODE}). The inset curves in red (top corners) show the spatiotemporal waveforms and those in black (bottom corners) are the frequency spectra. For both solitons and rolls the phases lie on straights lines of arbitrary slope (see Fig.~\ref{fig:prelim:numsol}). Parameter values are (a) $\alpha=2,\, d_2=-0.0124,\, F=1.41$, and (b) $\alpha=0,\, d_2=-0.0124,\, F=1.63$. The phase profile has been unwrapped in (b).}
\end{figure*}
\indent Numerical integration of Eq.~(\ref{eq:pl}) confirms the existence of the family of solutions found analytically. Our numerous runs of numerical integration, for different comb spans ($N$ from 3 to 1000) with random initial PDs taken from a uniform distribution over $(-\pi, \pi]$ always lead to PDs lying on straight lines. The slope of the line depends on the initial conditions. In Fig.~\ref{fig:prelim:numsol}, we show two examples, in blue (dark gray) and red (light gray), for a comb with 201 teeth and with two different sets of initial conditions. Figure~\ref{fig:prelim:numsol}(a) shows the initial conditions while Fig.~\ref{fig:prelim:numsol}(b) depicts the steady-state PDs at the end of the simulation time vs.\ mode number. The results shown in Fig.~\ref{fig:prelim:numsol} are for a triangular power spectrum given by $\mathord{a_\eta\propto\exp(-k_0|\eta|)}$, with $k_0 = 0.05$. This profile assumes a linear decay (in logarithmic scale) of the comb teeth power spectrum \cite{akhmediev2011triangular} with slope $\mathord{\propto-20k_0}$~dB per increasing mode number by unity. We found that the model is robust and addition of static randomness of modest relative size to the power spectrum and coupling coefficients $K(l,\eta)$ will still lead to aligned PDs. Also, through numerical integration of Eq.~(\ref{eq:pl}), we found that phase alignment occurs for a variety of power spectrum profiles so long as the powers of the sidebands are smaller than the pumped mode power. Additionally, we observe that specific features like cusp points or isolated sharp peaks in the power spectrum envelope lead to step-like signatures in the distribution of the steady-state PDs; this effect is a topic of ongoing investigations and will be reported elsewhere. \newline
\indent The phase alignment predicted by the reduced phase model of Eq.~(\ref{eq:pl}) is observed in the phase-locked solutions of the LLE. Figure~\ref{fig:prelim:phasealigned} shows two examples, in solitons and Turing rolls, where Eq.~(\ref{eq:LLE}) has been integrated numerically using the split-step Fourier transform method for a typical microresonator. In practice, the random initial phases arise from vacuum fluctuations that seed modulation instability or from the passage of the system through the chaotic state while changing the pump laser power or frequency. We note that the phase offset between the pumped mode and the rest of the phases emerges to counter dissipation \cite{loh2014phase, wen2014self}. \newline
\section{Stability of the Fixed Points}\label{sec:stability}
\begin{figure*}[htbp]
\centering
\includegraphics[width=0.7\textwidth]{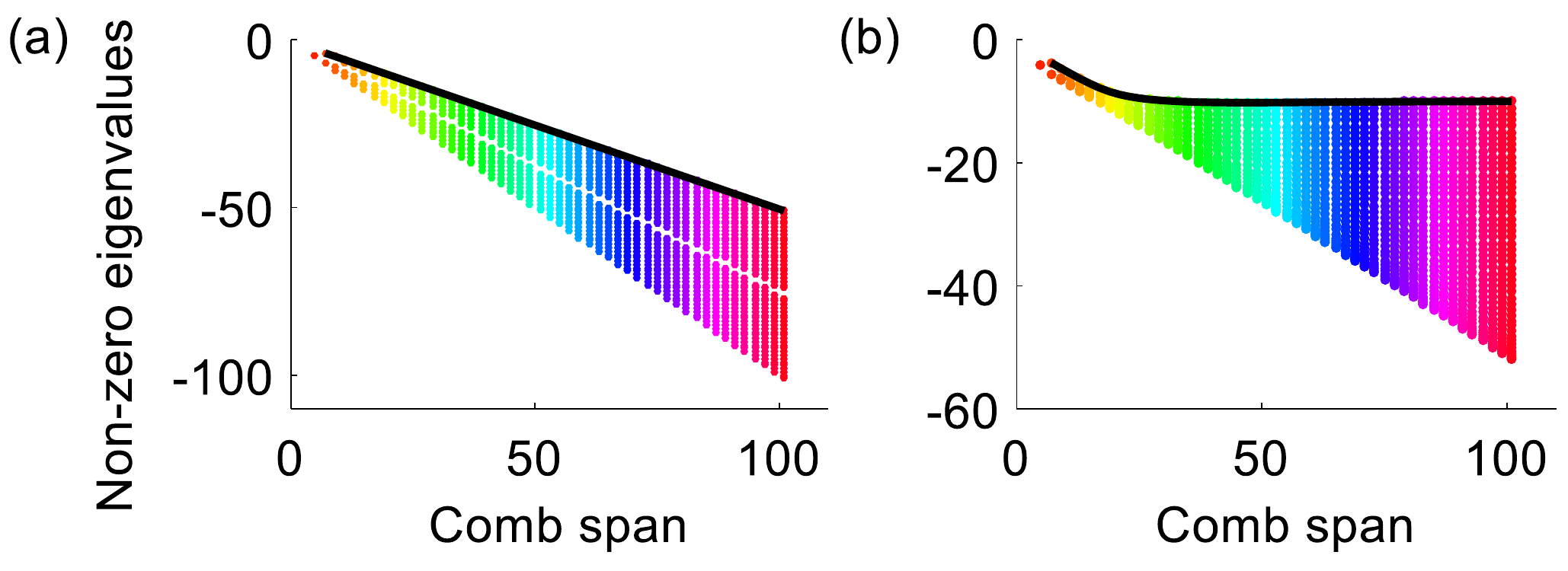}
\caption{\label{fig:prelim:eigs}Non-zero eigenvalues of the equilibrium (the Jacobian matrix $\bm{\mathrm{J}}$) vs.\ comb span for Eq.~(\ref{eq:pl}) for (a) uniform and (b) mode-number--dependent comb teeth magnitude profile of $\mathord{a_\eta\propto\exp(-k_0|\eta|)}$, ($k_0=0.1$). The $N$ distinct eigenvalues for each comb span are plotted with points of the same color excepting the negative eigenvalue of smallest magnitude, which is shown in black. The closest eigenvalues to zero for different spans form the black curves. Because these eigenvalues are dominant in determining stability, the black curves show that as the comb widens its stability improves for constant power spectrum, as seen in (a). For the realistic magnitude profile, on the other hand, the stability is not expected to improve, as is shown in (b). The stability of a steady-state solution of the LLE is, therefore, affected by its power profile. The effect of amplitudes can be taken into account through the coupling coefficients in Eq.~(\ref{eq:pl}).}
\end{figure*}

\noindent Next, we consider the linear stability of the solutions of Eq.~(\ref{eq:pl}). This analysis shows that the comb power spectrum profile significantly affects its stability properties \cite{chembo2014stability, parra2014dynamics}. We note that the analysis presented here is based on the reduced phase model and does not consider instabilities caused by comb power fluctuations. For the case of Turing patterns with multi-FSR spacing between adjacent comb teeth, in general cavity modal resonances not hosting comb power can also contribute to comb instability. However, because parametric gain for these modes is absent or small (depending on parametric gain bandwidth and the spectral distance of such modes from power-hosting modes), and since stronger comb teeth dominate the FWM process, such instabilities are less likely to grow. In fact, unless pump power and detuning values place the system close to the boundary of Turing roll and soliton existence regions in the power vs.\ detuning plane \cite{chembo2014stability, parra2014dynamics}, Turing rolls are monostable, in the sense that, unlike solitons, for the same system parameters and independent of system history or initial conditions only one Turing pattern with a unique number of peaks around the resonator will be realized \cite{pfeifle2015optimally}. \newline
\indent For simplicity, we take $k=0$. (Stability analysis for $k\ne0$ follows in a similar way.) We consider a frequency comb with $2N+1$ phase-locked teeth and temporarily ignore the dependence of the comb teeth magnitudes on mode number, i.e., as in \cite{wen2014self}, we take $a_{\eta\ne0}=a \ll a_0$. (The effect of the mode number dependence will be included shortly.) After phase locking, the centered phase averages $\zeta_\eta$ reach a steady-state value independent of mode number $\eta$ (since the phases $\phi_\eta$ lie on a straight line). Therefore, the coupling coefficients are all equal, i.e., $K(l,\eta)=K > 0$. Setting $\Delta_\eta = s_0 \eta + \epsilon_\eta$, we linearize Eq.~(\ref{eq:pl}) to get $\dot{\bm{\mathrm{\epsilon}}} = \bm{\mathrm{J}}\cdot \bm{\mathrm{\epsilon}}$, where $\bm{\mathrm{\epsilon}}=(\epsilon_1, \epsilon_2, \dots, \epsilon_N)^\mathrm{T}$, and the Jacobian $\bm{\mathrm{J}}$ and its eigenvalues can be expressed in closed form for any $N$ (see Appendix~\ref{app:stability}). Except for one zero eigenvalue, all of the eigenvalues are negative and real, indicating asymptotic stability of the synchronized state. The zero eigenvalue (corresponding to the Goldstone mode associated with the translational invariance of the system in the real space \cite{goldstone1962brokensym, anderson1984basicCMP}) is forced by the rotational symmetry of the LLE. In other words, the choice of origin for the azimuthal angle $\theta$ is arbitrary and leads merely to an added linear phase. This confirms the physical intuition that the slope of the phase profile of a soliton or Turing roll, determined by random initial conditions, is indeed arbitrary. Figure~\ref{fig:prelim:eigs}(a) shows the non-zero eigenvalues of the equilibrium for increasing comb span for the case of uniform sideband power profile. It is seen that the eigenvalue closest to zero (black curve) grows more negative with increasing comb span. The stability of the fixed points for each comb span is determined by the negative eigenvalue of smallest size. Hence, for the case of constant comb amplitudes, a wider comb is expected to demonstrate superior phase stability.\newline
\indent The model introduced in Eq.~(\ref{eq:pl}) allows the comparison of the phase stability properties of frequency combs with different power spectra. Because the coupling coefficients $K(l, \eta)$ depend on the comb teeth magnitudes, the power spectrum profile of a steady-state solution is expected to influence its stabiity. To investigate the effect of a non-constant comb power spectrum, we use a triangular comb power profile given by $\mathord{a_\eta\propto\exp(-k_0|\eta|)}$ \cite{akhmediev2011triangular}. Though not analytically tractable, we find numerically that again, except for a single zero eigenvalue forced by symmetry, the eigenvalues of $\bm{\mathrm{J}}$ all have negative real part. Figure~\ref{fig:prelim:eigs}(b) shows the eigenvalue spectrum vs.\ increasing comb span for the triangular power profile. Note that as the comb span increases, the eigenvalue of smallest magnitude becomes bounded and almost independent of $N$ (black curve in Fig.~\ref{fig:prelim:eigs}(b)). Therefore, the phase stability of the comb does not improve or degrade with increasing comb span when the natural mode number dependence of the comb teeth magnitudes is taken into account. Pfeifle \emph{et al.}\ \cite{pfeifle2015optimally} showed that in the presence of pump power and frequency noise, Turing rolls are more robust than solitons in the same microresonator with comparable pump powers. They attributed this finding partially to the smaller number of comb teeth in Turing rolls compared to solitons (see the Supplementary Material of \cite{pfeifle2015optimally}). The reduced phase model of Eq.~(\ref{eq:pl}) is derived with the assumption that there is a priori non-zero power in the comb teeth and therefore does not explicitly include the role of MI gain. Hence, our analysis here separates the influence of phase instabilities and shows that so far as \emph{phase} fluctuations are concerned, a smaller number of comb teeth does not enhance comb stability. Combined with the results of Ref.~\cite{pfeifle2015optimally}, this study suggests that MI gain and comb teeth power fluctuations significantly influence the stability of Turing rolls.\newline
\section{Discussion}\label{sec:discussion}
\noindent The existence of a self-synchronization mechanism explains soliton generation by both through-chaos \cite{coen2013universal, lamont2013route, herr2014temporal} and chaos-avoiding \cite{jaramillo2015deterministic} trajectories in the power-detuning plane. In either case, the parameter sweep creates a comb with single-FSR spacing. Sweeping through chaotic states provides a diverse pool of initial conditions which increases the odds of achieving phase-locked clusters (i.e., peaks) that subsequently grow into solitons; however, even without passing through chaos, the self-synchronization mechanism can generate solitons. It is worth noting that while we have focused on the phase-locked solutions of the LLE, this equation displays chaotic behavoir as well \cite{chembo2014stability, coillet2014routes}. Bifurcation to chaos in the reduced model of Eq.~(\ref{eq:pl}) can be understood through randomly oscillating coupling coefficients. While the model is robust and addition of static randomness of modest relative size to the coupling coefficients $K(l,\eta)$ will still lead to aligned PDs, our numerical simulations show that rapid random fluctuations of the comb teeth amplitudes (and therefore the coupling coefficients) hinder convergence of the phases toward a fixed point of the system. As a result, the phases will continue to wonder chaotically around without reaching a steady-state. Studying the behavior of this model in the presence of noise is an ongoing work and will be discussed elsewhere. \newline
\indent Phase measurements of stable optical frequency combs have shown that apart from combs with aligned phases (Fig.~\ref{fig:prelim:phasealigned}), phase spectra with $\pi$ and $\pi/2$ jumps can also arise in microcombs \cite{del2015phase}. We note that phase alignment governed by the reduced model is not contradictory to these phase jumps; combs with phase jumps have been constructed numerically as a sum of multiple solutions of the LLE (e.g., interleaved combs \cite{del2015phase} or solitons on an equally-spaced grid around the resonator with one solition removed or slightly shifted away from its location on the equidistant grid points \cite{lamb16stabilizing}) and their power spectra are more complicated than the smooth spectra of a Turing roll or soliton (as depicted in the insets in Fig.~\ref{fig:prelim:phasealigned}) considered in this work. It has been noted that avoided mode crossings \cite{herr2014mode} far from the pump are necessary for the experimental demonstration (through tuning the CW pump laser) and stabilization of such combs \cite{lamb16stabilizing, taheri2017optlattice}.\newline
\section{Summary and Outlook}\label{sec:summary}
\noindent In summary, we have introduced a reduced model for phase locking and the emergence of coherent spatiotemporal patterns in the damped, driven NLSE. This novel model underscores the fundamental link between spatiotemporal pulse formation and mode synchronization, and embodies the conservation of energy and momentum through ternary phase couplings. We have found attracting solutions of this phase model corresponding to dissipative solitons and Turing rolls and studied their stability, highlighting the significance of frequency comb power spectrum profile on it stability properties. \newline
\indent Although we have compared our results with micro-resonator-based optical frequency combs, they should apply to mode-locked laser systems as well. Gordon and Fisher's statistical mechanical theory describes the onset of laser pulsations as a first-order phase transition, treating the modes as the elementary degrees of freedom \cite{gordon2002phase}. Their ordered collective state is analogous to our synchronized dynamical attractor. The same controlling nonlinearity appears in both Eq.~(\ref{eq:pl}) and the master equation for passive mode locking based on a saturable absorber \cite{haus2000mode}, which approximates the absorber with a cubic nonlinearity \footnote{Comparison of Eq.~(16) in \cite{haus2000mode} with the LLE reveals their close similarity. In the LLE, there is an extra detuning term and the gain term is replaced by an external drive.}. We therefore expect the same dynamical mechanism to be responsible for the creation of sharp pulses in passively mode-locked lasers, despite different physical sources of optical gain (population inversion and stimulated emission versus parametric amplification). What matters is the fundamental link between spatiotemporal pulse formation and mode synchronization.

\begin{acknowledgments}
H.T. and K.W. thank Brian Kennedy and Andrey Matsko for many useful discussions. They also thank Rick Trebino for insightful discussions on mode locking in femtosecond lasers. K.W. thanks Henry Wen and Steve Strogatz for generously discussing the details of their results reported in \cite{wen2014self}. The authors thank one of the reviewers for constructive comments. H.T. was supported by the Air Force Office of Scientific Research Grant No.~2106DKP.
\end{acknowledgments}

\appendix
\section{Derivations}\label{app:derivation}
This Appendix details derivations leading to the equations in Sec.~(\ref{sec:reduction}).

The intra-cavity spatiotemporal field envelope $\psi(\theta, \tau)$ and the complex-valued comb teeth amplitudes $\tilde{a}_\eta$, $\mathord{\eta\in\{0, \pm1}, \pm2, \pm3, \dots\}$, are discrete-time Fourier transform pairs related through the following equations
\begin{equation}\label{eqn:FT_psi}
\psi(\theta, \tau)=\sum_{\eta=-\infty}^\infty \tilde{a}_\eta(\tau) \,  \mathrm{e}^{+\imi\eta\theta},
\end{equation}
and
\begin{equation}\label{eqn:FT_amps}
\tilde{a}_\eta(\tau)=\frac{1}{2\pi}\int_{-\pi}^{\pi}\difd\theta \, \psi(\theta, \tau) \, \mathrm{e}^{-\imi\eta\theta}.
\end{equation}
The summation in Eq.~(\ref{eqn:FT_psi}) is truncated and $\infty$ is replaced by the positive integer $N$ \cite{oppenheim1989dsp}. Using these equations and exploiting $\int_{-\pi}^\pi \, \mathrm{d}\theta \, \exp[\imi(\eta-\eta')\theta]=2\pi\delta_{\eta\eta'}$, it is straightforward to find the equivalent coupled nonlinear ordinary differential equations of Eq.~(\ref{eq:CNODE}) from the LLE. In the strong pumping regime and after using $\tilde{a}_\eta=a_\eta\exp(\imi\phi_\eta)$ in the nonlinear ODEs, the equations for the magnitudes $a_\eta$ and phases $\phi_\eta$ can be separated to yield
\begin{widetext}
\begin{align}\label{eq:amps}
\difdisp{}{\tau}\ln(a_\eta)&=-1+\frac{a_{-\eta}}{a_\eta}a_0^2\sin(\phi_\eta+\phi_{-\eta}-2\phi_0)+\frac{F_\mathrm{P}}{a_\eta}\cos(\phi_\mathrm{P}-\phi_\eta)\delta_{0\eta}\\ \nonumber
&-\frac{a_0}{a_\eta}\sum_l a_l \{2a_{\eta+l}\sin(\phi_0-\phi_l+\phi_{\eta+l}-\phi_\eta)+a_{\eta-l}\sin(\phi_l-\phi_0+\phi_{\eta-l}-\phi_\eta)\},
\end{align}
\begin{align}\label{eq:phases}
\difdisp{}{\tau}\phi_\eta &= 2a_0^2-\alpha+\frac{1}{2}d_2\eta^2+\frac{a_{-\eta}}{a_\eta}a_0^2\cos(2\phi_0-\phi_\eta-\phi_{-\eta})+\frac{F_\mathrm{P}}{a_\eta}\sin(\phi_\mathrm{P}-\phi_\eta)\delta_{0\eta} \\ \nonumber
&+\frac{a_0}{a_\eta}\sum_l a_l \{2a_{\eta+l}\cos(\phi_0-\phi_l+\phi_{\eta+l}-\phi_\eta)+a_{\eta-l}\cos(\phi_l-\phi_0+\phi_{\eta-l}-\phi_\eta)\}.
\end{align}
Using Eq.~(\ref{eq:phases}) and considering the symmetry of the power spectrum, the equations of motion for the centered phase averages $\zeta_\eta=(\phi_\eta+\phi_{-\eta})/2-\phi_0$ and phase differences $\Delta_\eta=(\phi_\eta-\phi_{-\eta})/2$ can be found,
\begin{align}\label{eq:averages}
\difdisp{}{\tau}\zeta_\eta &= \frac{1}{2}d_2\eta^2+a_0^2[1+\cos(2\zeta_\eta)]-\frac{F_\mathrm{P}}{a_\eta}\sin(\phi_\mathrm{P}-\phi_\eta)\delta_{0\eta} \\ \nonumber
& + \frac{a_0}{a_\eta}\sum_l a_l a_{\eta-l}\cos(\Delta_l+\Delta_{\eta-l}-\Delta_{\eta})\{2\cos(\zeta_{\eta-l}-\zeta_\eta-\zeta_l)+ \cos(\zeta_{\eta-l}-\zeta_\eta+\zeta_l)\},
\end{align}
\begin{align}\label{eq:PDs}
\difdisp{}{\tau}\Delta_\eta = \frac{a_0}{a_\eta}\sum_l a_l a_{\eta-l}\{2\sin(\zeta_{\eta-l}-\zeta_\eta-\zeta_l)+ \sin(\zeta_{\eta-l}-\zeta_\eta+\zeta_l)\}\sin(\Delta_l+\Delta_{\eta-l}-\Delta_{\eta}).
\end{align}
\end{widetext}
Equations~(\ref{eq:amps}, \ref{eq:phases}) for $\eta = 0$ lead to Eq.~(\ref{eq:pumpedamp}, \ref{eq:pumpedphase}) of the main text, and Eq.~(\ref{eq:PDs}) is the same as Eq.~(\ref{eq:pl}) in the main text, where the coupling coefficient $K(l, \eta)$ was defined. We note that the normalized chromatic dispersion coefficient $d_2$ is defined by $d_2 = -2D_2/\Delta\omega_0$, where $\Delta\omega_0$ is the linewidth of the pumped mode and $D_2$ is the second-order dispersion parameter found from the Taylor expansion of the cavity modal frequencies $\omega_\eta$ in the mode number $\eta$ at the pumped mode $\omega_0$ through $\omega_\eta = \omega_0 + D_1\eta + \frac{1}{2!}D_2\eta^2+\frac{1}{3!}D_3\eta^3+\dots$ . In the latter expression, $D_1$ is the resonator FSR (in rad/s) at the pumped mode.

To lowest non-zero order in $a_{\eta\ne 0}$, Eq.~(\ref{eq:averages}) becomes Eq.~(\ref{eq:avetrunc}). This equation is separable, i.e.,
$$
\int_{\zeta_\eta(\tau_0)}^{\zeta_\eta(\tau)} \frac{\mathrm{d}\zeta_\eta}{1+C(\eta)+\cos(2\zeta_\eta)}=a_0^2\int_{\tau_0}^{\tau}\mathrm{d}\tau',
$$
and can be integrated directly to give
\begin{equation}\label{eq:antisym_integrated}
\frac{1}{\sqrt{C(C+2)}}\tan^{-1}\left[\sqrt{\frac{C}{C+2}}\tan(\zeta_\eta)\right]_{\zeta_\eta(\tau_0)}^{\zeta_\eta(\tau)}=a_0^2(\tau-\tau_0).
\end{equation}
In these equations $C=d_2\eta^2/2a_0^2-F_\mathrm{P}\sin(\phi_\mathrm{P}-\phi_0)/a_0^3$, and $\tau_0$ accounts for constants of integration (or initial conditions). The latter equality can be written as
\begin{equation}\label{eq:antisym_tan}
\tan[\zeta_\eta(\tau)]=\sqrt{\frac{C+2}{C}}\tan[\sqrt{C(C+2)}a_0^2(\tau-\tau_0')],
\end{equation}
where $\tau_0'$ accounts for the constants of integration on both sides of Eq.~(\ref{eq:antisym_integrated}). The parameter $C$ appears in two combinations, $C/(C+2)$ and $C(C+2)$; if $-2<C<0$, then both expressions will be negative and the tangent on the right of Eq.~(\ref{eq:antisym_tan}) changes to a hyperbolic tangent. Therefore, one arrives at Eq.~(\ref{eq:antisym}) in the main text.

It is straightforward to show that the gain of modulation instability (MI) for the LLE of Eq.~(\ref{eq:LLE}) is given by \cite{chembo2010spectrum, chembo2014stability}
$$
\Gamma = \operatorname{Re}\left\lbrace{-1+\sqrt{a_0^4-\left(\alpha-\frac{1}{2}d_2\eta^2-2a_0^2\right)^2}}\right\rbrace,
$$
where $\operatorname{Re}\{\cdot\}$ denotes real part. For this expression to be positive, the following inequality should hold
\begin{equation}\label{eq:MIgain_cond}
a_0^4- 1 \ge \left( \alpha-\frac{1}{2}d_2\eta^2-2a_0^2\right)^2 \ge 0.
\end{equation}
It can simply be shown that the condition $-2<C<0$ on $C$ is equivalent to $a_0^2 \ge \abs{\alpha-d_2\eta^2/2-2a_0^2}$, which is guaranteed to hold in the presence of MI gain, cf. Eq.~(\ref{eq:MIgain_cond}). 

\section{Linear stability analysis}\label{app:stability}
In this Appendix, we review the stability analysis of the reduced phase model and introduce the generic form of the Jacobian matrix $\bm{\mathrm{J}}$ and its eigenvalues for the case of uniform comb amplitudes.

We consider Eq.~(\ref{eq:pl}) in the main text for a comb with $2N+1$ phase-locked teeth. For all the indices appearing in this equation to be in the range $[-N, N]$, the summation should run from $-(N-\eta)$ to $N$, i.e.,
$$
\difdisp{\Delta_\eta}{\tau}=a_0\sum\nolimits_{l=-(N-\eta)}^N K(l,\eta) \sin(\Delta_l+\Delta_{\eta-l}-\Delta_\eta).
$$
As explained in the main text, the coupling coefficients $K$ will be the same for uniform comb magnitude spectrum (where $a_{\eta\ne0}=a \ll a_0$).
\begin{widetext}
If each phase $\phi_\eta$ is perturbed from its steady-state value by $e_\eta$, the phase difference $\Delta_\eta=s_0\eta$ will change to $s_0\eta+\epsilon_\eta$, where $\epsilon_\eta=(e_\eta-e_{-\eta})/2$. Plugging into the above equation (Eq.~(\ref{eq:pl}) of the main text) and linearizing in $\epsilon_\eta$, we find the matrix equation $\dot{\bm{\mathrm{\epsilon}}} = \bm{\mathrm{J}}\cdot \bm{\mathrm{\epsilon}}$, for the perturbation vector $\bm{\mathrm{\epsilon}}=(\epsilon_1, \epsilon_2, \dots, \epsilon_N)^\mathrm{T}$. For $a_{\eta\ne0}=a \ll a_0$ the Jacobian $\bm{\mathrm{J}}$ and its eigenvalues can be expressed in closed form for any integer $N$. For $N$ an odd integer
$$
\bm{\mathrm{J}}=\begin{bmatrix}
-2N & 0 & \dots & 0 & 0 & 0 & \dots & 0 & 2 \\
0 & -2N+1 & \dots & \vdots & \vdots & \vdots & \dots & 2 & 2 \\
0 & 0 & \dots & 0 & 0 & 0 & \dots & 2 & 2 \\
0 & 0 & \dots & -2N-1+\floor{N/2} & 0 & 2 & \dots & 2 & 2 \\
\vdots & \vdots & \dots & 0 & -N+1-\floor{N/2} & 2 & \dots & \vdots & \vdots \\
0 & 0 & \dots & 2 & 2 & -N+2-\floor{N/2} & \dots & 2 & 2 \\
0 & 0 & \dots & 2 & 2 & 2 & \dots & 2 & 2 \\
0 & 2 & \dots & \vdots & \vdots & \vdots & \dots & -N & 2 \\
2 & 2 & \dots & 2 & 2 & 2 & \dots & 2 & -N+1
\end{bmatrix},
$$
and its eigenvalues are $0,-N-1,-N-2,\dots,-N+1-\floor{N/2},-N-\floor{N/2},-2N-2+\floor{N/2},-2N-3+\floor{N/2},\dots,-2N,-2N-1$ (where $\floor{\cdot}$ is the floor function). For even $N$, the Jacobian takes the following form
$$
\bm{\mathrm{J}}=\begin{bmatrix}
-2N & 0 & \dots & 0 & 0 & \dots & 0 & 2 \\
0 & -2N+1 & \dots & \vdots & \vdots & \dots & 2 & 2 \\
0 & 0 & \dots & 0 & 0 & \dots & 2 & 2 \\
\vdots & \vdots & \dots & -2N-1+N/2 & 2 & \dots & \vdots & \vdots \\
0 & 0 & \dots & 2 & -N+2-N/2 & \dots & 2 & 2 \\
0 & 0 & \dots & 2 & 2 & \dots & 2 & 2 \\
0 & 2 & \dots & \vdots & \vdots & \dots & -N & 2 \\
2 & 2 & \dots & 2 & 2 & \dots & 2 & -N+1
\end{bmatrix}.
$$
The eigenvalues of this matrix are $0, -N-1, -N-2, \dots, -N+2-N/2, -N+1-N/2, -2N-2+N/2, -2N-3+N/2, \dots, -2N, -2N-1$.
\end{widetext}
It is noted that there will always be a zero eigenvalue enforced by symmetry, and all other eigenvalues are negative. The negative eigenvalue of smallest size ($-N-1$) determines the stability of the fixed points. These eigenvalues for different comb spans ($2N+1$) are plotted in black in Fig.~\ref{fig:prelim:eigs}(a).


\bibliography{references}

\end{document}